\documentclass[onecolumn,usenatbib]{revtex4}
\usepackage{amsmath}
\usepackage[dvips]{graphicx}
\begin{document}

\title{Exact General Relativistic Discs and the Advance of Perihelion}

\author{D. Vogt\footnote{e-mail: dvogt@ime.unicamp.br}
 and 
P.S. Letelier\footnote{e-mail: letelier@ime.unicamp.br} }
\address{Departamento de Matem\'{a}tica Aplicada-IMECC, Universidade Estadual
de Campinas-UNICAMP,  13083-970 Campinas, S.\ P., Brazil}

\begin{abstract}
The advance of perihelion for geodesic motion on the galactic plane of some exact 
general relativistic disc solutions is calculated. Approximate analytical and numerical results 
are presented for the static Chazy-Curzon and the Schwarzschild discs in Weyl coordinates, the 
Schwarzschid disc in isotropic coordinantes and
the stationary Kerr disc in the Weyl-Lewis-Papapetrou metrics. It is found that for these disc 
models the advance of perihelion may be an increasing or decreasing function of the orbital 
excentricity. The precession due to Newtonian gravity for these disc models is also calculated. 
\end{abstract}

\maketitle
\section{Introduction}

The explanation of the anomalous precession of Mercury's orbit by Einstein in 1915 was one of the first 
successful predictions of General Relativity. In recent times the subject has received considerable attention with 
the possibility of high precision measurements of general relativistic effects in the orbits of binary pulsars 
like the PSR 1913+16 system discovered in 1974 \citep{h1}. 
\citet{s1} discuss in depth higher order general relativistic contributions 
to the periastron advance of binary pulsars. \citet{s2} calculated 
the periastron advance for a system composed of a Kerr black hole and an orbiting star. 

In the context of axially symmetric solutions of Einstein vacuum equations, 
\citet{b3} calculated the effect of different general relativistic multipole 
expansions in the advance of perihelion of test particles orbiting static axially symmetric attraction 
centres. \citet*{b2} derived approximate expressions for the periastron 
shift for motion in static and stationary axially symmetric spacetimes. However, we 
did not find in the literature similar works for general relativistic solutions with matter, in particular, 
disc-like configurations. Several such solutions can be found in the literature, e.g., 
\citet{m2}; \citet{b6}; \citet{l3}; \citet{c3}; \citet{l1}; \citet{b4}; \citet{b5}; \citet{l2}; 
\citet{n1}; \citet{p1}; \citet{g3}.   

The aim of this work is to study the advance of perihelion for motion of test particles in 
the galactic plane for a few exact solutions of Einstein field equations that represent 
disc-like configurations of matter (\citet{b1}; \citet{g1,g2}; \citet{v1,v2,v3}).
We derive approximate expressions and also present numerical results. We find that 
when matter is present the periastron shift may be an increasing or an decreasing function 
of the orbital excentricity.       
The paper is divided as follows: in Section \ref{sec_gen} we present the formalism to 
calculate the periastron shift 
for relativistic elliptic orbits of test particles. This formalism is then applied in 
Section \ref{sec_static} to two exact models of static relativistic discs in canonical Weyl 
coordinates and to one solution expressed in isotropic coordinates. In Section \ref{sec_station} 
we calculate the periastron shift for a solution of a rotating disc obtained from the 
Kerr metric. In Section \ref{sec_newt} we calculate the contribution of Newtonian 
gravity to the precession in the presented disc models so that it can be separated 
from relativistic effects. Finally, in Section \ref{sec_disc} we present a short discussion of the results. 
Along the work we take units such that $c=G=1$.   
\section{Advance of the Perihelion in Relativistic Orbits} \label{sec_gen}

In this section we derive the formulae to calculate the advance of perihelion for 
geodesic elliptic-like orbits in an axial symmetric spacetime with 
cylindrical coordinates $(t,r,z,\varphi)$. We follow closely \citet{b2}. Let us assume a test 
particle is bound in an elliptic orbit on the plane $z=0$. This orbit can be parametrized as
\begin{equation} \label{eq_orbit}
r=\frac{d(1-e^2)}{1+e\cos \chi} \mbox{,}
\end{equation} 
where $d$ and $e$ are, respectively, the ellipse's semi-major axis and excentricity, and 
$\chi$ is a variable called relativistic anomaly. From equation (\ref{eq_orbit}) we see that the minimum 
value $r_m=d(1-e)$ is obtained for $\chi=0$ and the maximum value $r_p=d(1+e)$ when $\chi=\pi$. 
At these points the equation $\mathrm{d}r/\mathrm{d}\varphi$ that describes the shape of the 
orbit vanishes. The relation between the functions $\varphi$ e $\chi$ can be expressed as
\begin{equation}
\frac{\mathrm{d}\varphi}{\mathrm{d}\chi}=\frac{ed(1-e^2)\sin \chi}{(1+e \cos \chi)^2}
\left. \frac{\mathrm{d}\varphi}{\mathrm{d}r} \right|_{r=r(\chi)} \mbox{,}
\end{equation}
where equation (\ref{eq_orbit}) was used. By symmetry, the change in the coordinate $\varphi$ 
when $\chi$ decreases from $\pi$ to $0$ is the same that when $\chi$ increses from 
$0$ to $\pi$; thus the total change in the coordinate $\varphi$ in one revolution is 
$2(\varphi(\pi)-\varphi(0))$, where
\begin{equation} \label{eq_prec}
\varphi(\pi)-\varphi(0)= \int_0^{\pi} \frac{ed(1-e^2)\sin \chi}{(1+e \cos \chi)^2}
\left. \frac{\mathrm{d}\varphi}{\mathrm{d}r} \right|_{r=r(\chi)} \mathrm{d} \chi \mbox{.}
\end{equation}
In a closed ellipse $\varphi$ would change by $2\pi$ per revolution, so the 
orbit precesses by an angle
\begin{equation} 
\Delta \varphi =2(\varphi(\pi)-\varphi(0)) -2\pi
\end{equation}
in one revolution. In general it is not possible to express the integral equation (\ref{eq_prec}) in terms 
of elementary functions; we will evaluate it numerically and also derive approximate expressions. 
\section{Advance of the Perihelion and Static Relativistic Discs} \label{sec_static}

We study first the precession of perihelion for orbits in static relativistic disc models in 
Weyl coordinates and isotropic coordinates. 
\subsection{Weyl Coordinates} \label{ss_weyl}

The general metric for a static axially symmetric spacetime in Weyl's canonical coordinates 
$(t,r,z,\varphi)$ is given by
\begin{equation} \label{eq_metric_w}
\mathrm{d}s^2=-e^{2\Phi}\mathrm{d}t^2+e^{-2\Phi}\left[ e^{2\Lambda}(\mathrm{d}r^2+\mathrm{d}z^2)+r^2\mathrm{d}\varphi^2 \right] \mbox{,}
\end{equation}
where $\Phi$ and $\Lambda$ are functions of $r$ and $z$. The Einstein vacuum equations 
for this metric reduce to the Weyl equations (\citealt{w1,w2})
\begin{gather}
\Phi_{,rr}+\frac{\Phi_r}{r}+\Phi_{,zz}=0 \mbox{,} \label{eq_weyl1}\\
\Lambda_r=r(\Phi_r^2-\Phi_z^2) \text{,} \qquad 
\Lambda_z=2r \Phi_r \Phi_z \mbox{.} \label{eq_weyl3}
\end{gather}
We shall consider two solutions of equations (\ref{eq_weyl1})--(\ref{eq_weyl3}): the Chazy-Curzon 
solution (\citealt{c1}; \citealt{c2})
\begin{equation} \label{eq_chazy}
e^{2\Phi}=e^{-2m/R} \text{,} \qquad e^{2\Lambda}=e^{-m^2r^2/R^4} \mbox{,}
\end{equation} 
where $R=\sqrt{r^2+z^2}$, and the Schwarzschild solution, expressed as (\citealt{w1})
\begin{equation} \label{eq_schw}
\Phi =\frac{1}{2} \ln \left[ \frac{R_1+R_2-2m}{R_1+R_2+2m} \right] \text{, } \quad
\Lambda =\frac{1}{2} \ln \left[ \frac{(R_1+R_2)^2-4m^2}{4R_1R_2} \right] \mbox{,}
\end{equation}
with $R_1^2=r^2+(m+z)^2$, $R_2^2=r^2+(-m+z)^2$. 

Let us briefly recall a procedure to generate disc-like distributions of matter given 
a vacuum solution of Einstein field equations. Mathematically, it consists in applying 
a transformation $z \rightarrow \mathsf{h}(z)+a$ on a given vacuum solution and then calculate the 
resulting energy-momentum tensor using Einstein's field equations. Thin discs can be obtained if we 
choose $\mathsf{h}=|z|$. For instance, \citet{b1} constructed thin discs 
using the Curzon solution equation (\ref{eq_chazy}) and the Schwarzschild solution equation (\ref{eq_schw}). 
On the other hand, thick discs can be constructed starting with the same vacuum solutions and using 
a class of even polynomials for $\mathsf{h}(z)$; see 
\citet{g2} and \citet{v2} for details. Also, a transformation originally proposed by \citet{m1}
 with $\mathsf{h}(z)=\sqrt{z^2+b^2}$ was used by \citet{v3} 
to generate relativistic disc-like distributions of matter from the Schwarzschild vacuum solution in 
isotropic coordinates. For our analysis, the advance of perihelion is always calculated on the 
galactic plane $z=0$,  where all the above mentioned transformations reduce to a constant. Thus, 
the results apply equally to thin and to thick discs. Henceforth this constant will be denoted $a$.  

For timelike orbits on the $z=0$ plane, the Lagrangean associated to metric equation (\ref{eq_metric_w}) 
reads
\begin{equation} \label{eq_lagrange}
2\mathcal{L}=-1=-e^{2\Phi}\dot{t}^2+e^{2(\Lambda-\Phi)}\dot{r}^2+r^2e^{-2\Phi}\dot{\varphi}^2 \mbox{,}
\end{equation}
where dots indicate differentiation with respect to proper time. Due to the independence of $\mathcal{L}$ 
from $t$ e $\varphi$, the conserved energy $E$ and angular momentum $h$ per unit mass can be
introduced
\begin{equation} \label{eq_const_w}
E= e^{2\Phi}\dot{t} \text{,} \qquad h= r^2e^{-2\Phi}\dot{\varphi} \mbox{.}
\end{equation} 
Using equation (\ref{eq_const_w}), the expression for the shape of the orbit follows from equation (\ref{eq_lagrange})
\begin{equation} \label{eq_orb_w}
\frac{\mathrm{d}r}{\mathrm{d}\varphi}=\frac{r}{e^{\Lambda}}\left[ 
\frac{r^2e^{-2\Phi}\left( E^2e^{-2\Phi}-1 \right)}{h^2}-1\right]^{1/2} \mbox{.}
\end{equation}
For an elliptic orbit with excentricity $e$ and semi-major axis $d$, the two constants of 
motion can be calculated by substituting $r_m=d(1-e)$ and $r_p=d(1+e)$ in 
$\mathrm{d}r/\mathrm{d}\varphi=0$ and solving the system. We have
\begin{equation} \label{eq_const2_w}
E^2=\frac{r_p^2e^{-2\Phi_p}- r_m^2e^{-2\Phi_m}}{r_p^2e^{-4\Phi_p}- r_m^2e^{-4\Phi_m}} 
\text{,} \qquad h^2=\frac{r_p^2r_m^2e^{-2(\Phi_p+\Phi_m)}(e^{-2\Phi_m}-e^{-2\Phi_p})}
{ r_p^2e^{-4\Phi_p}- r_m^2e^{-4\Phi_m}} \mbox{,}
\end{equation} 
where $\Phi_m=\Phi(r_m)$ and $\Phi_p=\Phi(r_p)$. 

To estimate the advance of perihelion for orbits in the $z=0$ plane for both disc models, 
it is reasonable to suppose that $m/d$ and $a/d$ are small quantities and expand equation (\ref{eq_prec}) 
with equations (\ref{eq_orb_w})--(\ref{eq_const2_w}) in multivariate Taylor series. For the Curzon 
disc, the expansion up to third order reads
\begin{equation} \label{eq_prec_cur}
\Delta \varphi = \frac{6\pi m}{d(1-e^2)}-\frac{3\pi a^2}{d^2(1-e^2)^2}+\frac{\pi m^2(44-9e^2)}{
2d^2(1-e^2)^2}-\frac{6\pi ma^2(6+e^2)}{d^3(1-e^2)^3}+\frac{\pi m^3(192-53e^2)}{2d^3(1-e^2)^3} \mbox{,}
\end{equation}
 while for the Schwarzschild disc we obtain
 \begin{equation} \label{eq_prec_sch} 
\Delta \varphi = \frac{6\pi m}{d(1-e^2)}-\frac{3\pi a^2}{d^2(1-e^2)^2}+\frac{3\pi m^2(14-3e^2)}{
2d^2(1-e^2)^2}-\frac{6\pi ma^2(6+e^2)}{d^3(1-e^2)^3}+\frac{3\pi m^3(56-19e^2)}{2d^3(1-e^2)^3} \mbox{.}
\end{equation}
Both expansions with $a=0$ agree with those presented by \citet{b2} up to second order. 
In equations (\ref{eq_prec_cur})--(\ref{eq_prec_sch}) the terms corresponding to the vacuum 
solutions are all positive, whereas the ones related to the presence of matter (parameter 
$a$) have negative signs. When matter is absent, the angle of advance is an increasing 
function with respect to excentricity, but expansions (\ref{eq_prec_cur})--(\ref{eq_prec_sch}) 
suggest that this may not be true in the present disc models. By imposing $\partial (\Delta \varphi)/\partial e=0$ in 
equations (\ref{eq_prec_cur})--(\ref{eq_prec_sch}), the 
following expressions for the parameter $a$ are obtained, respectively,
\begin{gather} 
a^2=\frac{m\left[ 12d^2(1-e^2)^2+md(1-e^2)(79-9e^2)+m^2(523-106e^2)\right]}{12\left[ d(1-
e^2)+m(19+2e^2)
\right]} \mbox{,} \label{eq_est_curz}\\
a^2=\frac{m\left[ 4d^2(1-e^2)^2+md(1-e^2)(25-3e^2)+m^2(149-38e^2)\right]}{4\left[ d(1-
e^2)+m(19+2e^2)
\right]} \mbox{,} \label{eq_est_sch}
\end{gather}
These expressions, evaluated for $e=0$ and $e=1$, yield the results
\begin{gather}
a^2=\frac{m(12d^2+79md+523m^2)}{12(d+19m)} \text{, } \quad a^2=\frac{139m^2}{84}
\quad
\text{ (Curzon)}\mbox{,} \label{eq_est_curz2} \\
a^2=\frac{m(4d^2+25md+149m^2)}{4(d+19m)} \text{, } \quad a^2=\frac{37m^2}{28}  \quad
\text{ (Schw.)}\mbox{.}   \label{eq_est_sch2}
\end{gather}
Equations (\ref{eq_est_curz2})--(\ref{eq_est_sch2}) give an estimate for the ranges 
in the parameter $a$ for which the angle of advance as a function of excentricity 
has a critical point.    
\begin{figure}
\centering
\includegraphics[scale=0.675]{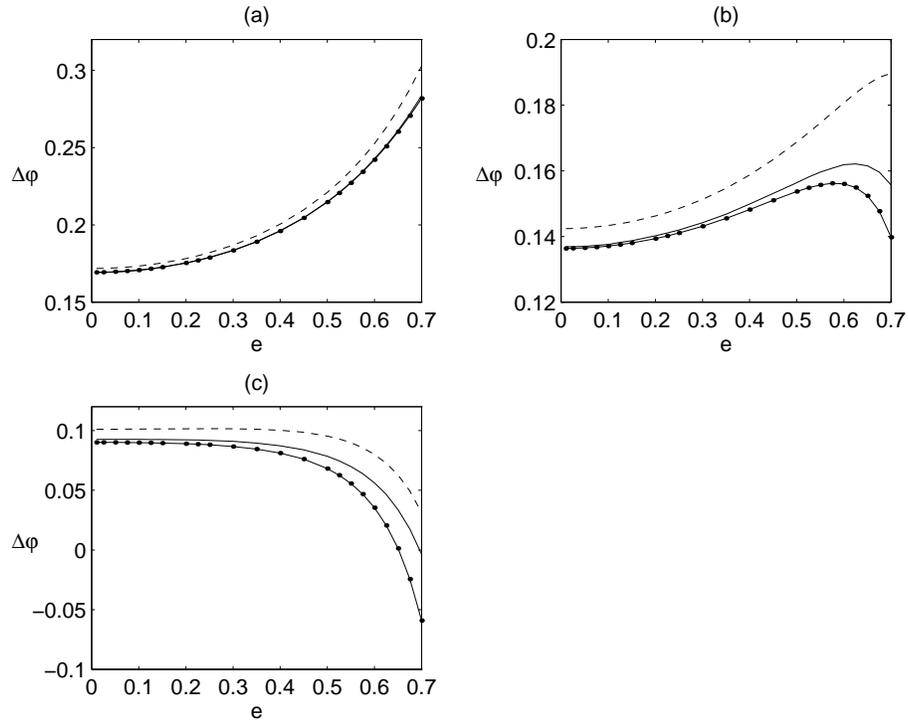}
\caption{The advance of the perihelion $\Delta \varphi$ as function of 
excentricity $e$ for the Chazy-Curzon disc. Parameters: $m/d=0.01$, $a/d=0.05$ 
in (a), $a/d=0.075$ in (b) and $a/d=0.1$ in (c).
Solid lines: numerical integration of equation (\ref{eq_prec}). Dashed lines:
values obtained from equation (\ref{eq_prec_cur}) up to terms of second order. 
Dotted lines: the same expansion with terms of third order.} \label{fig_1}
\end{figure}
\begin{figure}
\centering
\includegraphics[scale=0.675]{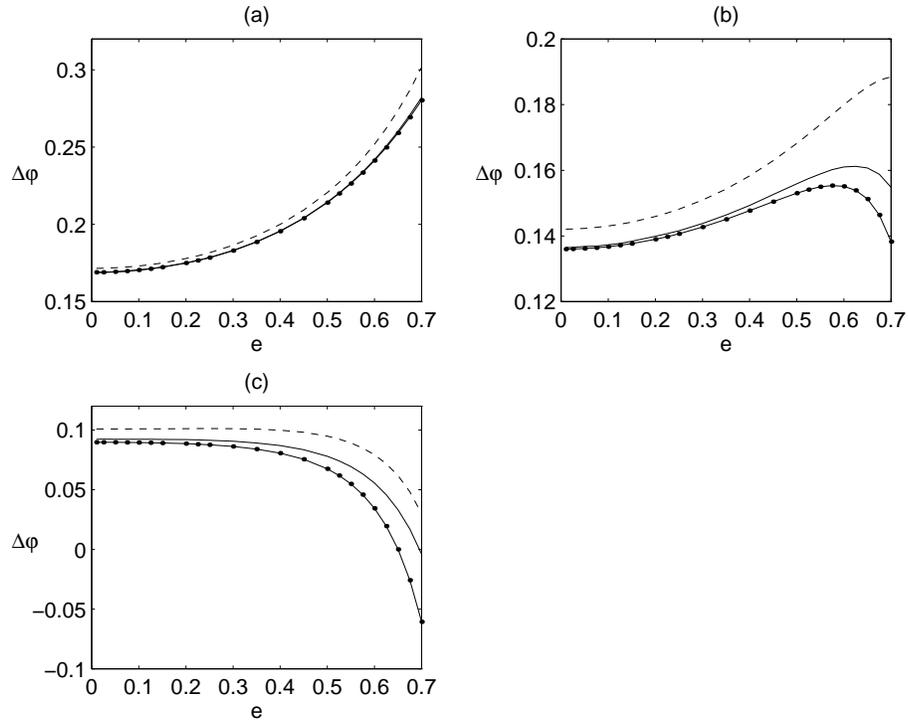}
\caption{The perihelion shift $\Delta \varphi$ as function of 
excentricity $e$ for the Schwarzschild disc. Parameters: $m/d=0.01$, $a/d=0.05$ 
in (a), $a/d=0.075$ in (b) and $a/d=0.1$ in (c).
Solid lines: numerical integration of equation (\ref{eq_prec}). Dashed lines:
values obtained from equation (\ref{eq_prec_sch}) up to terms of second order. 
Dotted lines: the same expansion with terms of third order.} \label{fig_2}
\end{figure}

Figs \ref{fig_1}(a)--(c) show the angle of precession as function of 
excentricity for the Chazy-Curzon disc with parameters $m/d=0.01$, $a/d=0.05$ 
in fig.\ \ref{fig_1}(a), $a/d=0.075$ in (b) and $a/d=0.1$ in (c). The curves 
with solid lines were calculated by numerical integration of equation (\ref{eq_prec}), 
those with dashed lines represent expansion equation (\ref{eq_prec_cur}) with terms 
up to second order, and the dotted lines the same expression but with terms 
of third order. Figs \ref{fig_2}(a)--(c) show the results for the Schwarzschild 
disc with the same value of parameters. In both cases the numerical results and 
the expansion up to third order agree well for small values of $e$. Using the 
estimates given by equations (\ref{eq_est_curz2}) and (\ref{eq_est_sch2}), the 
intervals with critical points would be $0.0128 \leq a/d \leq 0.0948$ and 
$0.0115 \leq a/d \leq 0.0946$, respectively. 
\subsection{Isotropic Coordinates} \label{ss_iso}

The line element in isotropic form in cylindrical coordinates $(t,r,z,\varphi)$ 
may be expressed as
\begin{equation} \label{eq_metric_iso}
\mathrm{d}s^2=-e^{2\Phi}\mathrm{d}t^2+e^{2\Lambda}(\mathrm{d}r^2+\mathrm{d}z^2+
r^2\mathrm{d}\varphi^2) \mbox{,}
\end{equation}
where the $\Phi$ and $\Lambda$ are only functions of $r$ and $z$. The vacuum 
Schwarzschild solution for metric equation (\ref{eq_metric_iso}) has the form
\begin{equation} \label{eq_sch_i}
e^{2\Phi}=\left( \frac{1-\frac{m}{2R}}{1+\frac{m}{2R}} \right)^2 \text{,} \qquad
e^{2\Lambda}= \left( 1+\frac{m}{2R} \right)^4 \mbox{,}
\end{equation}
where $R=\sqrt{r^2+z^2}$. Also here disc-like distributions of matter can be 
generated by applying convenient transformations on the $z$ coordinate (see, for example
 \citealt{g2}; \citealt{v1,v2,v3}), as was discussed in Section \ref{ss_weyl}. 

For metric equation (\ref{eq_metric_iso}), the shape of the orbit of a test-particle 
confined on the $z=0$ plane is described by
\begin{equation} \label{eq_orb_i}
\frac{\mathrm{d}r}{\mathrm{d}\varphi}=r \left[ \frac{r^2e^{2\Lambda}\left( E^2e^{-2\Phi}-1
\right)}
{h^2}-1 \right]^{1/2} \mbox{.}
\end{equation}
The constants of motion $E$ and $h$ are now given by the following expressions
\begin{equation} \label{eq_const2_i}
E^2=\frac{r_p^2e^{2\Lambda_p}-r_m^2e^{2\Lambda_m}}{r_p^2e^{2(\Lambda_p-\Phi_p)}-
r_m^2
e^{2(\Lambda_m-\Phi_m)}} \text{,} \qquad
h^2=\frac{r_p^2r_m^2e^{2(\Lambda_p+\Lambda_m)}
(e^{-2\Phi_m}-e^{-2\Phi_p})}{r_p^2e^{2(\Lambda_p-\Phi_p)}-r_m^2e^{2(\Lambda_m-
\Phi_m)}}
\mbox{,}
\end{equation}
with the same notation as defined in Section \ref{ss_weyl}.

For the Schwarzschild disc in isotropic coordinates, an approximate expression 
for the precession of perihelion reads
\begin{equation} \label{eq_prec_sch_i}
\Delta \varphi = \frac{6\pi m}{d(1-e^2)}-\frac{3\pi a^2}{d^2(1-e^2)^2}+\frac{3\pi m^2(14-3e^2)}{
2d^2(1-e^2)^2}-\frac{6\pi ma^2(6+e^2)}{d^3(1-e^2)^3}+\frac{3\pi m^3(57-16e^2)}{2d^3(1-e^2)^3} \mbox{.}
\end{equation}
Comparing equations (\ref{eq_prec_sch}) and (\ref{eq_prec_sch_i}) reveals that they are 
almost identical, the difference beginning only in the last term. The calculation of
$\partial (\Delta \varphi)/\partial e=0$ provides
\begin{equation}
a^2=\frac{m\left[ 4d^2(1-e^2)^2+md(1-e^2)(25-3e^2)+m^2(155-32e^2)\right]}{4\left[ d(1-
e^2)+m(19+2e^2)
\right]} \mbox{,}
\end{equation}
which evaluated on $e=0$ and $e=1$ gives, respectively
\begin{equation} \label{eq_est_sch_i}
a^2=\frac{m(4d^2+25md+155m^2)}{4(d+19m)} \text{, } \quad a^2=\frac{41m^2}{28} \mbox{.}
\end{equation}
\begin{figure}
\centering
\includegraphics[scale=0.675]{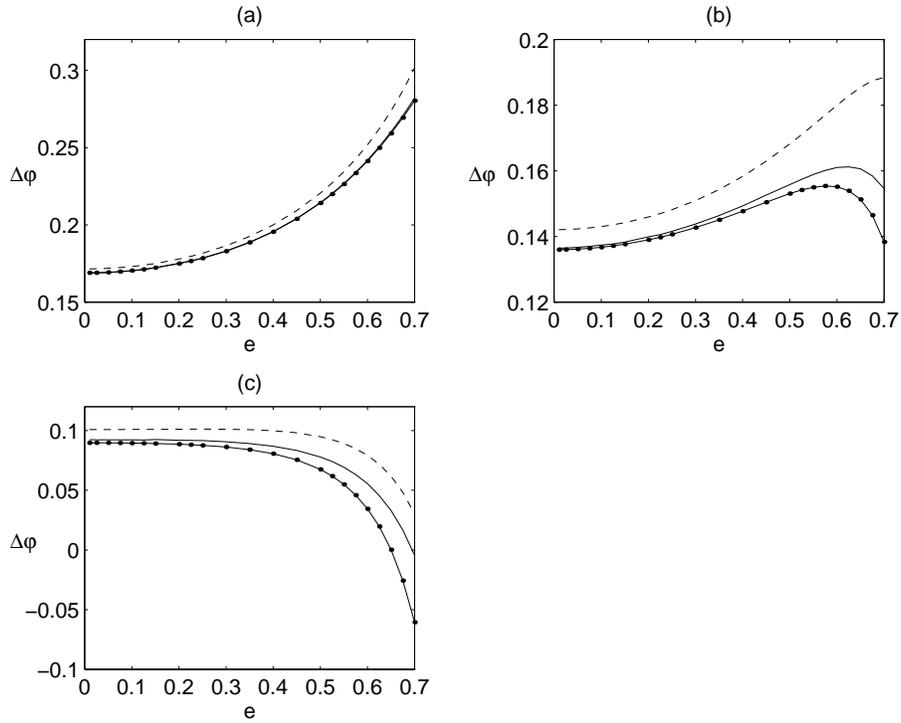}
\caption{The advance of the perihelion $\Delta \varphi$ as function of 
excentricity $e$ for the Schwarzschild disc in isotropic coordinates. Parameters: $m/d=0.01$, $a/d=0.05$ 
in (a), $a/d=0.075$ in (b) and $a/d=0.1$ in (c).
Solid lines: numerical integration of equation (\ref{eq_prec}). Dashed lines:
values obtained from equation (\ref{eq_prec_sch_i}) up to terms of second order. 
Dotted lines: the same expansion with terms of third order.} \label{fig_3}
\end{figure}
\begin{figure}
\centering
\includegraphics[scale=0.675]{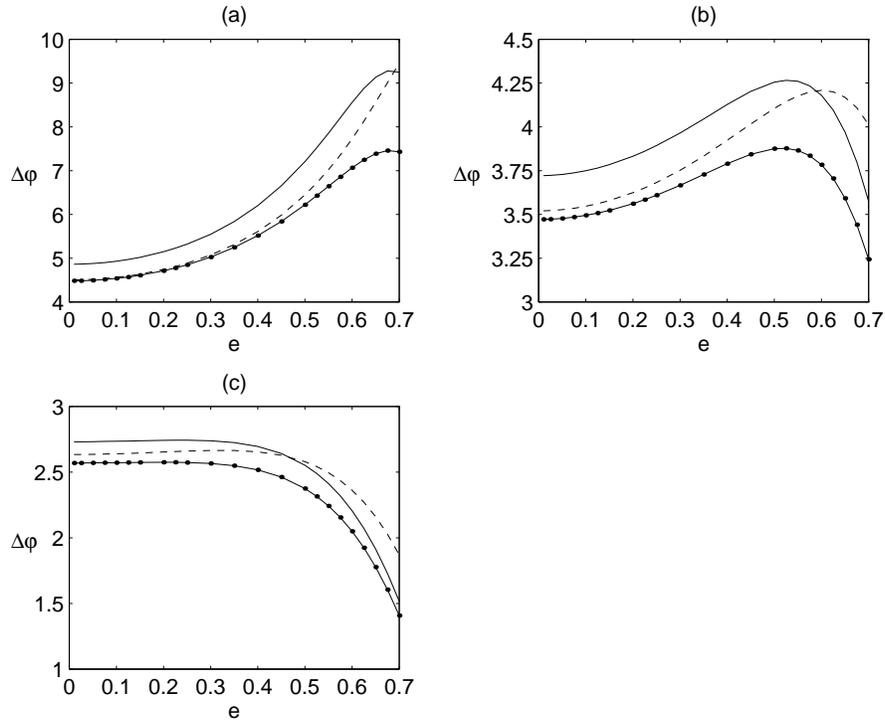}
\caption{The advance of the perihelion $\Delta \varphi$ as function of 
excentricity $e$ for the three disc models. Parameters: $m/d=0.15$, $a/d=0.175$ in (a), $a/d=0.225$ in
(b) and $a/d=0.275$ in (c). Solid lines: Chazy-Curzon disc. Dashed lines: Schwarzschild 
disc in Weyl coordinates. Dotted lines: Schwarzschild disc in isotropic coordinates.} \label{fig_4}
\end{figure}

Figs \ref{fig_3}(a)--(c) display some curves of the angle $\Delta \varphi$ as 
function of the excentricity $e$ for the Schwarzschild disc in isotropic 
coordinates with parameters $m/d=0.01$, $a/d=0.05$ in fig.\ \ref{fig_3}(a),
$a/d=0.075$ in (b) and $a/d=0.1$ in (c). The curves 
with solid lines were calculated by numerical integration of equation (\ref{eq_prec}), 
those with dashed lines represent expansion equation (\ref{eq_prec_sch_i}) with terms 
up to second order, and the dotted lines the same expression but with terms 
of third order. For this example, equation (\ref{eq_est_sch_i}) estimate an interval 
$0.0121 \leq a/d \leq 0.0947$. Finally in figs \ref{fig_4}(a)--(c) the three disc 
models are compared. Parameters were taken $m/d=0.15$ e $a/d=0.175$ in fig.\ \ref{fig_4}(a),
$a/d=0.225$ in (b) and $a/d=0.275$ in (c). All curves were obtained by 
numerical integration of equation (\ref{eq_prec}). Solid lines represent the results 
for the Chazy-Curzon disc, dashed lines for the Schwarzschild disc in Weyl coordinates 
and dotted lines for the Schwarzschild disc in isotropic coordinates. Remember that 
as the parameter $a$ is increased, all the discs become less relativistic 
(\citealt{b1}; \citealt{v1}). This is reflected in the numerical values of the precession 
angle, which are greater in fig.\ \ref{fig_4}(a) than in (c). Qualitatively the curves
for the three disc models are similar. As suggested by the expansions 
equations (\ref{eq_prec_sch}) and (\ref{eq_prec_sch_i}), both models obtained from 
the Schwarzschild solution give quite similar results for low excentric orbits.      
\section{Advance of the Perihelion and Stationary Relativistic Discs} \label{sec_station}

In this section we investigate the effect of rotation on the perihelion shift for 
stationary relativistic discs, in particular discs generated from the vacuum Kerr metric. We 
begin with the metric for a stationary axially symmetric spacetime 
\begin{equation} \label{eq_lewis}
\mathrm{d}s^2=-e^{2\Phi}\left( \mathrm{d}t+\mathcal{A}\mathrm{d}\varphi \right)^2+e^{-2\Phi}\left[ 
r^2\mathrm{d}\varphi^2+e^{2\Lambda} \left( \mathrm{d}r^2+\mathrm{d}z^2 \right)\right] \mbox{,}
\end{equation}
where $\Phi$, $\Lambda$ and $\mathcal{A}$ are functions of $r$ and $z$. The vacuum Kerr solution 
for metric equation (\ref{eq_lewis}) may be written as 
\begin{gather}
\Phi =\frac{1}{2}\ln \left[ \frac{(R_1+R_2)^2-4m^2+\alpha^2(R_1-R_2)^2/\sigma^2}
{(R_1+R_2+2m)^2+\alpha^2(R_1-R_2)^2/\sigma^2}\right] \mbox{,} \label{eq_kerr1}\\
\Lambda =\frac{1}{2}\ln \left[ \frac{(R_1+R_2)^2-4m^2+\alpha^2(R_1-R_2)^2/\sigma^2}{4R_1R_2}\right] \mbox{,} \\
\mathcal{A}=\frac{\alpha m}{\sigma^2}\frac{(R_1+R_2+2m)\left[ 4\sigma^2-(R_1-R_2)^2\right]}
{(R_1+R_2)^2-4m^2+\alpha^2(R_1-R_2)^2/\sigma^2} \mbox{,} \label{eq_kerr2}
\end{gather}
where $m$ and $\alpha$ are, respectively, the mass and the Kerr parameter, 
$R_1=\sqrt{r^2+(z+\sigma)^2}$, $R_2=\sqrt{r^2+(z-\sigma)^2}$ and $\sigma=\sqrt{m^2-\alpha^2}$.
Following the same procedure taken in Section \ref{sec_static}, the orbit's shape of a test particle on the 
plane $z=0$ for metric eq.\ (\ref{eq_lewis}) is described by
\begin{equation} \label{eq_orb_lewis}
\frac{\mathrm{d}r}{\mathrm{d}\varphi}=\frac{r}{e^{\Lambda}}\left[ 
\frac{r^2e^{-2\Phi}\left( E^2e^{-2\Phi}-1 \right)}{\left( E\mathcal{A}+h \right)^2}-1\right]^{1/2} \mbox{.}
\end{equation} 
The conserved energy $E$ and angular momentum $h$ are found by solving the system of equations 
\begin{gather}
r_p^2e^{-2\Phi_p}\left( E^2e^{-2\Phi_p}-1 \right)-\left( E\mathcal{A}_p+h \right)^2=0 \mbox{,} \\
r_m^2e^{-2\Phi_m}\left( E^2e^{-2\Phi_m}-1 \right)-\left( E\mathcal{A}_m+h \right)^2=0 \mbox{,}
\end{gather}
where $\mathcal{A}_p=\mathcal{A}(r_p)$, $\mathcal{A}_m=\mathcal{A}(r_m)$, and where again 
the same notation was used as defined in Section \ref{ss_weyl}. For a given excentricity $e$ and semi-major 
axis $d$ the system of equations admit two distinct solutions, corresponding to prograde $(h>0)$ and 
retrograde $(h<0)$ orbits. After applying a convenient transformation on the Kerr solution 
equations (\ref{eq_kerr1})--(\ref{eq_kerr2}) to generate stationary disc-like distributions of matter (see \citealt{g1}; \citealt{v4}), 
we assume the ratios $m/d$, $a/d$ e $\alpha/d$ to be small and expand equation (\ref{eq_prec}) in series up to 
third order\footnote{There is a discrepancy of a factor of 2 between our result and the second term of 
equation (34) in \citet{b2}. We compared both expressions with the numerical integration of the 
 exact expressions for vacuum and our result is closer to the numerical values.}
 \begin{multline} \label{eq_prec_ker}
\Delta \varphi = \frac{6\pi m}{d(1-e^2)} \mp \frac{8\pi \alpha m^{1/2}}{d^{3/2}(1-e^2)^{3/2}}
-\frac{3\pi a^2}{d^2(1-e^2)^2} +\frac{3\pi m^2(14-3e^2)}{2d^2(1-e^2)^2}
+\frac{3\pi \alpha^2}{d^2(1-e^2)^2} \mp \frac{12\pi \alpha m^{3/2}(5-e^2)}{d^{5/2}(1-e^2)^{5/2}} \\
-\frac{6\pi ma^2(6+e^2)}{d^3(1-e^2)^3}+\frac{3\pi m^3(56-19e^2)}{2d^3(1-e^2)^3} 
+\frac{6\pi m\alpha^2(12-e^2)}{d^3(1-e^2)^3} \mbox{,}
\end{multline}
where the minus (plus) sign refers to prograde (retrograde) orbits. The solution of 
$\partial (\Delta \varphi)/\partial e=0$ yields
\begin{multline} \label{eq_est_ker}
a^2=\frac{1}{4\left[ d(1-e^2)+m(19+2e^2) \right]} \left\{ m\left[ 4d^2(1-e^2)^2+md(1-e^2)(25-3e^2)
+m^2(149-38e^2)+4\alpha^2(35-2e^2)\right] \right. \\
\left. +4\alpha d^{1/2}(1-e^2)^{1/2} \left[ \alpha d^{1/2}(1-e^2)^{1/2} 
\mp 2m^{1/2}d(1-e^2)\mp m^{3/2}(23-3e^2) \right] \right\} \mbox{,}
\end{multline}
which for $e=0$ and $e=1$ simplifies to
\begin{gather}
a^2=\frac{m(4d^2+25md+149m^2+140\alpha^2)+4\alpha d^{1/2}(\alpha d^{1/2} \mp 2m^{1/2}d
\mp 23m^{3/2})}{4(d+19m)} \mbox{,} \label{eq_est2_ker} \\
a^2=\frac{37m^2+44\alpha^2}{28} \mbox{,} \label{eq_est3_ker}
\end{gather}  
 respectively. 
 
\begin{figure}
\centering
\includegraphics[scale=0.675]{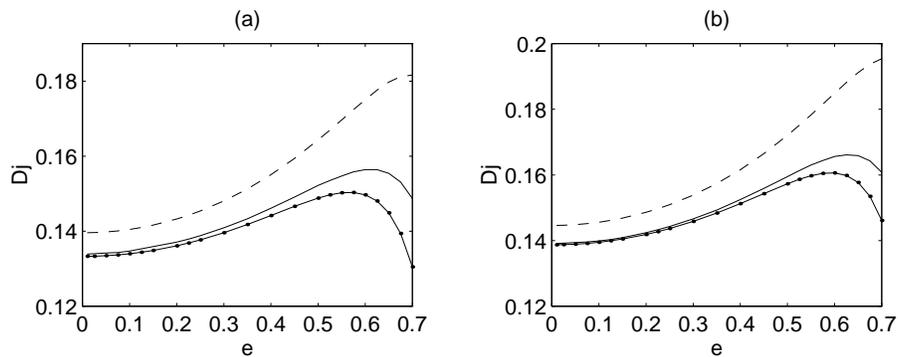}
\caption{The perihelion shift $\Delta \varphi$ as function of 
excentricity $e$ for the Kerr disc. Parameters: $m/d=0.01$, $a/d=0.075$ and 
$\alpha/d=0.001$. Fig. \ref{fig_5}(a) displays results for prograde orbits and 
(b) for retrograde orbits. 
Solid lines: numerical integration of equation (\ref{eq_prec}). Dashed lines:
values obtained from equation (\ref{eq_prec_ker}) up to terms of second order. 
Dotted lines: the same expansion with terms of third order.} \label{fig_5}
\end{figure}
\begin{figure}
\centering
\includegraphics[scale=0.65]{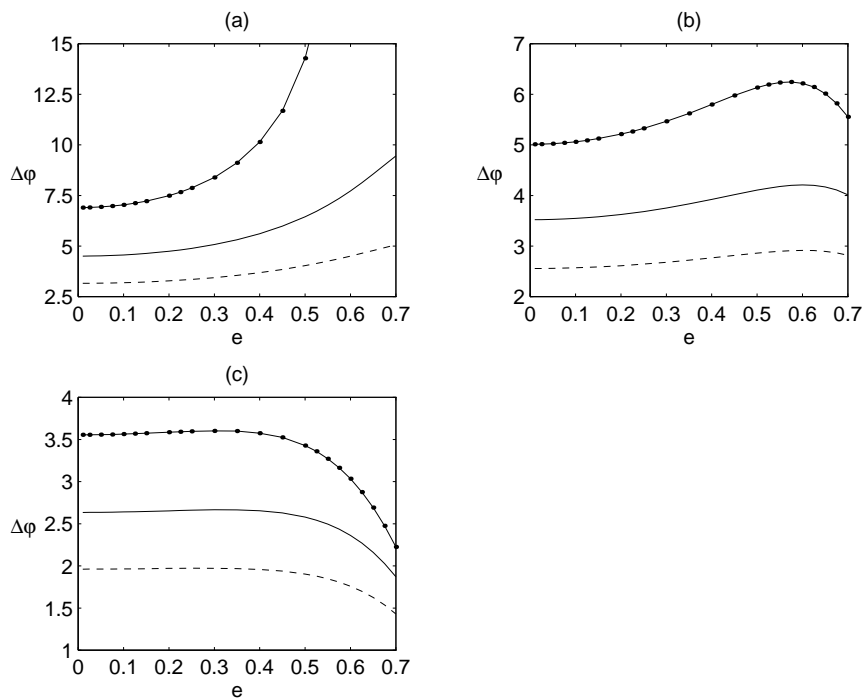}
\caption{The perihelion shift $\Delta \varphi$ as function of 
excentricity $e$ for the Kerr disc. Par\^ameters: $m/d=0.15$, $\alpha/d=0.05$, $a/d=0.175$ in (a), 
$a/d=0.225$ in (b) and $a/d=0.275$ in (c). Solid lines: disc without rotation. Dashed lines: prograde 
orbits. Dotted lines: retrograde orbits.} \label{fig_6}
\end{figure}
Figs \ref{fig_5}(a)--(b) show the advance of the perihelion as function of excentricity for the Kerr 
disc with parameters $m/d=0.01$, $a/d=0.075$ and $\alpha/d=0.001$. The curves in (a) are the 
results for prograde orbits and (b) for retrograde orbits. Curves with solid lines were calculated 
by numerical integration of equation (\ref{eq_prec}), 
those with dashed lines represent expansion equation (\ref{eq_prec_ker}) with terms 
up to second order, and the dotted lines the same expression but with terms 
of third order. For these values we obtain from equation (\ref{eq_est3_ker}) intervals 
$0.0116 \leq a/d \leq 0.0970$ and $0.0116 \leq a/d \leq 0.0972$ for prograde and 
retrograde orbits, respectively. The results of numerical integration of the exact expressions 
are depicted in figs \ref{fig_6}(a)--(c) with parameter values $m/d=0.15$, $\alpha/d=0.05$, 
$a/d=0.175$ in (a), $a/d=0.225$ in (b) and $a/d=0.275$ in (c). The values for prograde orbits are 
represented by dashed lines, retrograde orbits by dotted lines, and solid lines the perihelion shift 
without rotation (Weyl disc). We note that for prograde orbits the Kerr parameter lowers the 
angle of precession and has an opposite effect for retrograde orbits. The signs in the expansion 
equation (\ref{eq_prec_ker}) also predict these effects. 

 \begin{figure}
\centering
\includegraphics[scale=0.65]{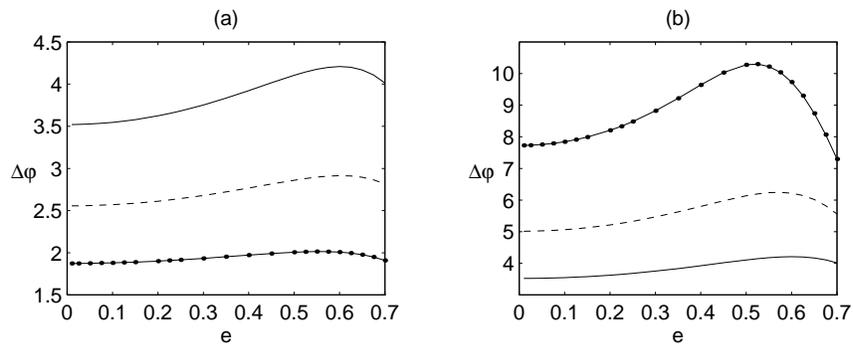}
\caption{The perihelion shift $\Delta \varphi$ as function of 
excentricity $e$ for the Kerr disc. In (a) are shown prograde orbits and in (b) retrograde orbits. 
Parameters: $m/d=0.15$, $a/d=0.225$, $\alpha/d=0$ (solid lines), $\alpha/d=0.05$ 
(dashed lines) and $\alpha/d=0.1$ (dotted lines).} \label{fig_7}
\end{figure} 
Finally in figs \ref{fig_7}(a)--(b) the parameters $m/d=0.15$ and $a/d=0.225$ were held constant and 
the Kerr parameter was changed: $\alpha/d=0$ (solid lines), $\alpha/d=0.05$ 
(dashed lines) and $\alpha/d=0.1$ (dotted lines). Prograde orbits are shown in fig.\ \ref{fig_7}(a) and 
retrograde orbits in (b). Rotation has the same effect on both types of orbits as observed in 
fig.\ \ref{fig_6}.
\section{Effect of Newtonian Precession} \label{sec_newt}

The advance of perihelion calculated for the relativistic models of discs presented in Sections 
\ref{sec_static} and \ref{sec_station} has two origins: one due to relativity and other from purely
Newtonian gravity, since any flattened body will generate a perihelion precession. Thus, it 
would be interesting to separate the relativistic from the Newtonian contributions to the 
precession.  

It can be shown that in the non-relativistic limit the above mentioned disc models reduce on the $z=0$ plane 
to the Kuzmin model \citep{k1}
\begin{equation} \label{eq_phi_n}
\Phi=-\frac{m}{\sqrt{r^2+a^2}} \mbox{.}
\end{equation}  
The orbital equation in the usual cylindrical coordinates reads 
\begin{equation}  \label{eq_orb_newt}
\frac{\mathrm{d}r}{\mathrm{d}\varphi}=r \left[ \frac{2r^2(E_M-\Phi)}{h^2}-1\right]^{1/2} \mbox{,}
\end{equation}
where $E_M$ is the conserved mechanical energy of the test particle
\begin{equation} \label{eq_const_n}
E_M=\frac{r_p^2\Phi_p-r_m^2\Phi_m}{r_p^2-r_m^2}\text{, and} \qquad 
h=\frac{2r_p^2r_m^2(\Phi_p-\Phi_m)}{r_p^2-r_m^2} \mbox{,}
\end{equation}
whith the notation as defined in Section \ref{ss_weyl}.
Proceeding as in the previous Sections, it is possible to deduce approximate expressions for 
the perihelion shift. Using equations (\ref{eq_prec}) and (\ref{eq_phi_n})--(\ref{eq_const_n}), 
one has the following expansion 
\begin{equation} \label{eq_prec_newt}
\Delta \varphi = -\frac{3\pi a^2}{d^2(1-e^2)^2}+\mathcal{O}((a/d)^4) \mbox{.} 
\end{equation}
Note that the angle of advance is independent of $m$, which is canceled in the fraction in 
equation (\ref{eq_orb_newt}). Thus, the first term of equation (\ref{eq_prec_newt}), which also appears 
in expansions (\ref{eq_prec_cur}), (\ref{eq_prec_sch}), (\ref{eq_prec_sch_i}) and (\ref{eq_prec_ker}), is the purely 
Newtonian contribution up to third order to the perihelion shift. 

 \begin{figure}
\centering
\includegraphics[scale=0.65]{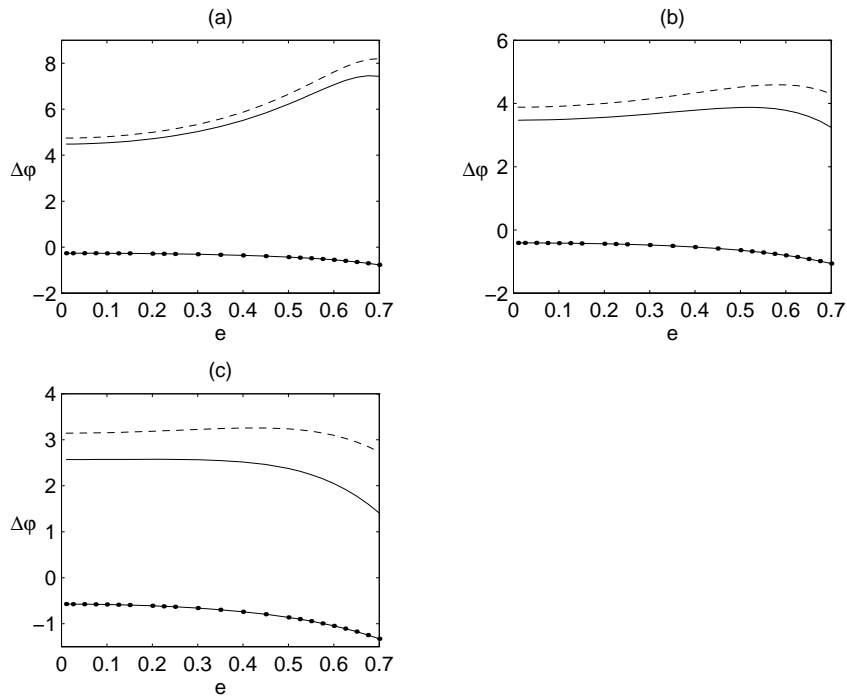}
\caption{The different contributions to the perihelion shift $\Delta \varphi$ as function of 
excentricity $e$ for the Schwarzschild disc in isotropic coordinates. Parameters: $m/d=0.15$, 
$a/d=0.175$ in (a), $a/d=0.225$ in (b) and $a/d=0.275$ in (c). 
Solid lines: the total shift. Dotted lines: the purely Newtonian contribution.
Dashed lines: the relativistic contribution.} \label{fig_8}
\end{figure}
As a numerical example, in figs \ref{fig_8}(a)--(c) we display curves of $\Delta \varphi$ as 
function of the excentricity $e$ for the Schwarzschild disc in isotropic coordinates 
with $m/d=0.01$, $a/d=0.175$ in (a), $a/d=0.225$ in (b) and $a/d=0.275$ in (c); the 
same values that were used in fig.\ \ref{fig_4}. The solid lines represent the total angle 
of precession; the dotted lines represent the precession due to Newtonian effects obtained
from the numerical integration of equation (\ref{eq_prec}) with equations 
(\ref{eq_phi_n})--(\ref{eq_const_n}), and the dashed lines are the difference between the 
previous two values. It is seen that the precession rate due to Newtonian gravity is in 
the opposite sense to the relativistic precession. Also for less relativistic discs the 
Newtonian contribution is more significant, as expected.

\section{Discussion} \label{sec_disc}

We studied the advance of perihelion for elliptic orbits of test particles in geodesic motion 
on the galactic plane for relativistic static and stationary disc models. We derived approximate 
expressions for the perihelion shift and compared them with the numerical integration of the 
exact solutions. The results show that the angle of advance can increase as well as decrease 
with increasing excentricity. We have that highly relativistic discs favours the first situation, and 
the advance of perihelion decreases with excentricity when the discs become less relativistic. 
The effect of rotation was also studied for a particular stationary disc model based on the Kerr 
solution. We found that the Kerr parameter lowers the perihelion shift for prograde orbits and 
increases it in the case of retrograde orbits. We also calculated the different contributions  (Newtonian and relativistic) to the  advance of perihelion for the relativistic disc models.

Our conclusions are based on the study of a few exact general relativistic disc models,
in particular, the  Miyamoto-Nagai model, that presents some characteristics of real galaxies. 
We believe that the results found may be common to other models of galaxies.

\section*{Acknowledgments}
 We thank FAPESP for financial support, P.\ S.\ L.\ also thanks CNPq.


\begin{thebibliography}{}
\bibitem[\protect\citeauthoryear{Bardeen \& Wagoner}{1971}]{b6} Bardeen J. M., Wagoner R. V., 1971, ApJ, 167, 359
\bibitem[\protect\citeauthoryear{Bi\v{c}\'{a}k \& Ledvinka}{1993}]{b4} Bi\v{c}\'{a}k J., Ledvinka T., 1993, Phys. Rev. Lett., 71, 1669
\bibitem[\protect\citeauthoryear{Bi\v{c}\'{a}k, Lynden-Bell \& Katz}{Bi\v{c}\'{a}k et al.}{1993}]{b1} Bi\v{c}\'{a}k J., 
Lynden-Bell D., Katz J., 1993, Phys. Rev. D, 47, 4334
\bibitem[\protect\citeauthoryear{Bi\v{c}\'{a}k, Lynden-Bell \& Pichon}{1993}]{b5} Bi\v{c}\'{a}k J., 
Lynden-Bell D., Pichon C., 1993, MNRAS, 265, 126
\bibitem[\protect\citeauthoryear{Bini, De Paolis, Geralico, Ingrosso \& Nucita}{Bini et al.}{2005}]{b2} Bini D., 
De Paolis F., Geralico A., Ingrosso G., Nucita A., 2005, Gen. Relativ. Gravit., 37, 1263 
\bibitem[\protect\citeauthoryear{Boisseau \& Letelier}{2002}]{b3} Boisseau B., Letelier P. S., 2002, Gen. Relativ. Gravit., 34, 1077
\bibitem[\protect\citeauthoryear{Chamorro, Gregory \& Stewart}{1987}]{c3} Chamorro A., Gregory R., Stewart J. M., 1987, 
Proc. R. Soc. London A, 413, 251. 
\bibitem[\protect\citeauthoryear{Chazy}{1924}]{c1} Chazy M., 1924, Bull. Soc. Math. France, 52, 17
\bibitem[\protect\citeauthoryear{Curzon}{1924}]{c2} Curzon H., 1924, Proc. London Math. Soc., 23, 477
\bibitem[\protect\citeauthoryear{Gonz\'{a}lez \& Espitia}{2003}]{g3} Gonz\'{a}lez G., Espitia O. A., 2003, Phys. Rev. D, 68, 104028
\bibitem[\protect\citeauthoryear{Gonz\'{a}lez \& Letelier}{2000}]{g1} Gonz\'{a}lez G., Letelier P. S., 
2000, Phys. Rev. D, 62, 064025
\bibitem[\protect\citeauthoryear{Gonz\'{a}lez \& Letelier}{2004}]{g2} Gonz\'{a}lez G., Letelier P. S., 
2004, Phys. Rev. D, 69, 044013
\bibitem[\protect\citeauthoryear{Hulse \& Taylor}{1975}]{h1} Hulse R. A., Taylor J. H., 1975, ApJ, 195, L51
\bibitem[\protect\citeauthoryear{Kuzmin}{1956}]{k1} Kuzmin G. G., 1956, Astron. Zh., 33, 27
\bibitem[\protect\citeauthoryear{Lemos}{1989}]{l1} Lemos J. P. S., 1989, Class. Quantum Grav., 6, 1219
\bibitem[\protect\citeauthoryear{Lemos \& Letelier}{1994}]{l2} Lemos J. P. S., Letelier P. S., 1994, Phys. Rev. D, 49, 5135
\bibitem[\protect\citeauthoryear{Lynden-Bell \& Pineault}{1978}]{l3} Lynden-Bell D., Pineault S., 1978, MNRAS, 185, 679
\bibitem[\protect\citeauthoryear{Miyamoto \& Nagai}{1975}]{m1} Miyamoto M., Nagai R., 1975, PASJ, 27, 533
\bibitem[\protect\citeauthoryear{Morgan \& Morgan}{1969}]{m2} Morgan T., Morgan L., 1969, Phys. Rev., 183, 1097
\bibitem[\protect\citeauthoryear{Neugebauer \& Meinel}{1995}]{n1} Neugebauer G., Meinel R., 1995, Phys. Rev. Lett., 75, 3046
\bibitem[\protect\citeauthoryear{Pichon \& Lynden-Bell}{1996}]{p1} Pichon C., Lynden-Bell D., 1996, MNRAS, 280, 1007
\bibitem[\protect\citeauthoryear{Sch\"afer \& Darmour}{1988}]{s1} Sch\"afer G., Darmour T., 1988, Nuovo Cimento B, 101(2), 127
\bibitem[\protect\citeauthoryear{Sch\"afer \& Wex}{1993}]{s2} Sch\"afer G., Wex N., 1993, Phys. Lett. A, 174, 196
\bibitem[\protect\citeauthoryear{Vogt \& Letelier}{2003}]{v1} Vogt D., Letelier P. S., 2003, Phys. Rev. D, 68, 084010
\bibitem[\protect\citeauthoryear{Vogt \& Letelier}{2005a}]{v2} Vogt D., Letelier P. S., 2005a, Phys. Rev. D, 71, 084030 
\bibitem[\protect\citeauthoryear{Vogt \& Letelier}{2005b}]{v3} Vogt D., Letelier P. S., 2005b, MNRAS, 363, 268
\bibitem[\protect\citeauthoryear{Vogt \& Letelier}{2007}]{v4} Vogt D., Letelier P. S., 2007, Phys. Rev. D, 76, 084010 
\bibitem[\protect\citeauthoryear{Weyl}{1917}]{w1} Weyl H., 1917, Ann. Phys., 54, 117
\bibitem[\protect\citeauthoryear{Weyl}{1919}]{w2} Weyl H., 1919, Ann. Phys., 59, 185
\end{thebibliography}
\end{document}